\documentclass[letterpaper, 10 pt, conference]{ieeeconf}  

\IEEEoverridecommandlockouts                              
\usepackage{graphics} 
\usepackage{epsfig} 
\usepackage{mathptmx} 
\usepackage{times} 
\usepackage{amsmath} 
\usepackage{amssymb}  
\usepackage{amsbsy}
\usepackage{xcolor}
\usepackage{booktabs}
\usepackage{caption}
\usepackage{tabulary}
\usepackage{tabularx}

\usepackage{algorithm}
\usepackage[noend]{algpseudocode}

\makeatletter
\def\BState{\State\hskip-\ALG@thistlm}
\makeatother

\newcommand*\xbar[1]{%
  \hbox{%
    \vbox{%
      \hrule height 0.5pt 
      \kern0.5ex
      \hbox{%
        \kern-0.1em
        \ensuremath{#1}%
        \kern-0.1em
      }%
    }%
  }%
} 

\newtheorem{rem}{Remark}

\title{\LARGE \bf
Online Identification of Time-Varying Systems: a Bayesian approach}

\author{G. Prando, D. Romeres and A. Chiuso $^\dagger$
\thanks{This work has been partially supported by the FIRB project ``Learning meets time'' (RBFR12M3AC) funded by MIUR.}
\thanks{$^\dagger$ Dept. of Information  Engineering, University of Padova (e-mail: \{\tt \small romeresd,prandogi,chiuso\}@dei.unipd.it)}%
}

\begin{document}

\maketitle
\thispagestyle{empty}
\pagestyle{empty}

\begin{abstract}
We extend the recently introduced regularization/Bayesian System Identification procedures to the estimation of time-varying systems. Specifically, we consider an online setting, in which new data become available at given time steps. The real-time estimation requirements imposed by this setting are met by estimating the hyper-parameters through just one gradient step in the marginal likelihood maximization and by exploiting the closed-form availability of the impulse response estimate (when Gaussian prior and Gaussian measurement noise are postulated). By relying on the use of a forgetting factor, we propose two methods to tackle the tracking of time-varying systems. In one of them, the forgetting factor is estimated by treating it as a hyper-parameter of the Bayesian inference procedure.
\end{abstract}

\section{Introduction}
\noindent The identification of time-varying systems plays a key role in different applications, such as adaptive and model predictive control, where a good real-time tracking of the system to be controlled is necessary. In addition, the detection of changes or drifts in plant parameters is crucial in terms of process monitoring and fault detection. Online System Identification (SysId) and the estimation of time-varying systems are typically strictly connected problems: one would like to exploit the new data that become available in order to track possible changes in the system dynamics.\\
Recursive Prediction Error Method (RPEM), a variant of the classical PEM \cite{Ljung:99,RecursiveBook}, represents nowadays a well-established technique, through which the current estimate can be efficiently updated, as soon as new data are provided. RPEM are parametric approaches, relying on Recursive Least-Squares (RLS) routines, which compute the parameter estimate by minimizing a functional of the prediction errors \cite{Ljung:99}. An extension of these approaches to the identification of time-varying systems involves the adoption of a forgetting factor, through which old data become less relevant in the estimation criterion. Convergence and stability properties of Forgetting Factor RPEM  have been well-studied within the SysId community \cite{bittanti1990convergence,guo1993performance}. \\
Alternative approaches model the coefficients trajectories as stochastic processes \cite{Chow84}, thus exploiting Kalman filtering \cite{guo1990estimating} or Bayesian inference \cite{Sarris73} for parameter estimation. Combinations of bases sequences (e.g. wavelet basis \cite{tsatsanis1993time}) have also been considered to model the parameters time evolution.
\\The above-mentioned parametric procedures share the criticality of the model selection complexity: this step is especially crucial when the model complexity has to be modified in response to changes in the true system dynamics.
In addition, classical complexity selection rules (e.g. cross-validation or information criteria) may not be applicable in online settings, due to the excessive computational effort they require. 
Model complexity issues have been partially addressed in the SysId community through the recent introduction of non-parametric methods, relying on Gaussian processes and Bayesian inference \cite{GP-AC-GdN:11,SurveyKBsysid}. In this framework model complexity is tuned in a continuous manner by estimating the hyper-parameters which describe the prior distribution chosen by the user \cite{PCAuto2015}. This property makes these new  techniques appealing for the online identification of time-varying systems: indeed, model complexity can be continuously adapted whenever new data become available.\\
In a previous work \cite{RPPCECC2016} we started exploring this research direction by adapting the newly introduced Bayesian procedures to an online identification setting. The methodologies proposed in \cite{RPPCECC2016} are extended in this new paper by dealing with time-varying systems. Two approaches, relying on the use of a forgetting factor, are proposed; in particular, following the approach in \cite{PA2014}, we investigate the online estimation of the forgetting factor by treating it as a hyper-parameter of the Bayesian inference procedure. These techniques are experimentally compared with the classical parametric counterparts: the results appear favourable and promising  for the methods we propose.
\\
The paper is organized as follows. Sec.~\ref{sec:problem_formulation} presents the online identification framework and the challenges we will try to address. Sec.~\ref{sec:pem} provides a brief review of parametric real-time identification techniques, while Sec.~\ref{sec:bayes} illustrates the Bayesian approach to linear SysId, both in the batch and online scenarios. In particular, Sec.~\ref{sec:time_var} focuses on the estimation of time-varying systems. Experimental results are reported in Sec.~\ref{sec:experiment}, while conclusions are drawn in Sec.~\ref{sec:conclusion}.

\section{Problem Formulation}\label{sec:problem_formulation}
\noindent Consider a dynamical system described through an output-error model, i.e.:
\begin{equation} \label{equ:sys}
y(t) = \left[h \ast u\right](t) + e(t), \quad y(t), \ u(t) \in\mathbb{R}
\end{equation}
where $h(t)$ denotes the model impulse response and $e(t)$ is assumed to be a zero-mean Gaussian noise with variance $\sigma^2$.\\
SysId techniques aim at estimating the impulse response $h$ of the system, once a set $\mathcal{D}:=\left\{y(t),u(t)\right\}_{t=1}^N$ of measurements of its input and output signals is provided.\\
In this work we consider an online setting, in which a new set of input-output measurements becomes available every $T$ time steps.
Specifically, let us define the variable $i:=k/T$ by assuming w.l.o.g. that $k$ is a multiple of $T$, and the $i^{th}-$dataset as $\mathcal{D}_i =\left\{u(t),y(t)\right\}_{t=(i-1)T +1}^{iT}$.
\\We suppose that at time $k$ an impulse response estimate $\hat{h}^{(i)}$ has been computed using the data coming from a collection of previous datasets $\bigcup_{l=1}^{i} \mathcal{D}_l = \left\{u(t),y(t)\right\}_{t=1}^{i T}$; at time $k+T$ new data $\mathcal{D}_{i+1}$ become available and we would like to update the previous estimate $\hat{h}^{(i)}$ by exploiting them. In addition we assume that the underlying system undergoes certain variations that we would like to track: this situation could often arise in practice, due to e.g. variations of the internal temperature, of the masses (e.g. after grasping an object).\\
Furthermore, online applications typically require that the new estimate is available before the new dataset $\mathcal{D}_{i+2}$ is provided, thus limiting the computational complexity and the memory storage of the adopted estimation methods.\\
In this paper, the recently proposed Bayesian approach to SysId \cite{SurveyKBsysid} is adapted in order to cope with the outlined online setting. Its performances are compared with the ones achieved using classical parametric approaches.

\begin{rem}
We stress that in the remainder of the paper we will use the indexes $k$ and $iT$ interchangeably.
\end{rem}

\section{Parametric Approach}\label{sec:pem}
\noindent Standard parametric approaches to SysId rely on the a-priori choice of a model class $\mathcal{M}$ (e.g. ARX, ARMAX, OE, etc.), which is completely characterized by a parameter $\theta \in \mathbb{R}^m$.

\subsection{Batch Approach}
\noindent In the batch setting, when a dataset $\mathcal{D}=	\left\{y(t),u(t)\right\}_{t=1}^N$ is provided, the identification procedure reduces to estimate $\theta$ by minimizing the sum of squared prediction errors:
\begin{equation}
\hat{\theta} = \arg\min_{\theta\in \mathbb{R}^m} V_N (\theta,\mathcal{D}) = \arg\min_{\theta\in \mathbb{R}^m}  \frac{1}{2}\sum_{t=1}^N \left(y(t)-\hat{y}(t\vert \theta)\right)^2
\end{equation}
where $\hat{y}(t\vert \theta)$ denotes the one-step ahead predictor \cite{Ljung:99}.

\subsection{Online Approach}
\noindent The extension of these procedures to an online setting relies on RLS (or pseudo LS) methods.\\
For ease of notation, let us assume $T=1$ in this section. Suppose that at time $k+1$ a new input-output data pair $\mathcal{D}_{i+1}$ is provided; then $\hat{\theta}^{(i)}$ is updated as:
\begin{equation}\label{equ:param_update}
\hat{\theta}^{(i+1)} = \hat{\theta}^{(i)} + \mu^{(i+1)} Q^{(i+1)^{-1}} \nabla_{\theta} V_{k+1}(\hat{\theta}^{(i)}, \textstyle{\bigcup_{l=1}^{i+1}}\mathcal{D}_{l})
\end{equation}
where $\nabla_\theta V_{k+1}(\hat{\theta}^{(i)}, \bigcup_{l=1}^{i+1}\ \mathcal{D}_{l})$ denotes the gradient of the loss function computed in the previous estimate and in the new data; $\mu^{(i+1)}\in \mathbb{R}$ and $Q^{(i+1)}\in\mathbb{R}^{m\times m}$ are appropriate scalings which assume different shapes according to the specific algorithm which is adopted (see \cite{RecursiveBook} and \cite{Ljung:99}, Ch. 11, for further details). Notice that \eqref{equ:param_update} is simply a scaled gradient step w.r.t. the loss function $V_{k+1}(\theta,\bigcup_{l=1}^{i+1}\mathcal{D}_{l})$.

\subsection{Dealing with time-varying systems}
\noindent In order to cope with time-varying systems, a possible strategy involves the inclusion of a \textit{forgetting factor} $\bar{\gamma}$ in the loss function $V_{k}(\theta,\mathcal{D})$:
\begin{equation} \label{equ:loss_ff}
V_{k}^\gamma (\theta,\mathcal{D}) =  \frac{1}{2} \sum_{t=1}^k \bar{\gamma}^{k-t} \left(y(t)-\hat{y}(t\vert \theta)\right)^2, \qquad \bar{\gamma} \in (0,1]
\end{equation}
In this way old measurements become less relevant for the computation of the estimate. A recursive update of the estimate $\hat{\theta}^{(i)}$ (as the one in \eqref{equ:param_update}) can be derived (\cite{Ljung:99}, Ch. 11).
\\
As an alternative, a sliding window approach can be adopted: at each time step only the last $N_w$ data are used for computing the current estimate (with $N_w$ being the window length). However, since this approach does not admit an update rule as the one in \eqref{equ:param_update}, the computational complexity of the new estimate will depend on the window length.
\\
A crucial role in the application of parametric SysId techniques is played by the model order selection step: once a model class $\mathcal{M}$ is fixed, its complexity has to be chosen using the available data. This is typically accomplished by estimating models with different complexities and by applying tools such as cross-validation or information criteria to select the most appropriate one. However, the estimation of multiple models may be computationally expensive, making this procedure not suited for the online identification of time-varying systems. Indeed, in this framework, it should ideally be applied every time new data become available.
\\
The recently proposed approach to SysId, relying on regularization/Bayesian techniques, overpasses the above-described issue by jointly performing estimation and order selection.
Next section will illustrate how the batch regularization/Bayesian method can be tailored to the online identification of time-varying systems.

\section{Regularization/Bayesian Approach} \label{sec:bayes}

\subsection{Batch Approach}
\label{subsec:batch_approach}
\noindent We discuss how the regularization/Bayesian technique works in the standard batch setting, i.e. when data $\mathcal{D}=\left\{y(t),u(t)\right\}_{t=1}^N$ are given. For future use, let us define the vector $Y_N =\left[y(1)\ ...\ y(N)\right]^\top\in\mathbb{R}^N$.
\\
According to the Bayesian estimation, the impulse response $h$ is considered as a realization of a stochastic process with a prior distribution $p_\eta(h)$, depending on some parameters $\eta\in \Omega$. The prior $p_\eta(h)$ is designed in order to account for some desired properties of the estimated impulse response, such as smoothness and stability \cite{GP-AC-GdN:11,SurveyKBsysid}. In the Bayesian framework, the parameters $\eta$ are known as hyper-parameters and they need to be estimated from the data, e.g. by optimizing the so-called marginal likelihood (i.e. the likelihood once the latent variable $h$ has been integrated out) \cite{PCAuto2015}:
\begin{equation}
\hat{\eta} =\arg\max_{\eta\in\Omega} p_\eta(Y_N) = \arg\max_{\eta\in\Omega}  \int p(Y_N\vert h)p_\eta(h)dh
\end{equation}
Once the hyper-parameters $\eta$ have been estimated, the minimum variance estimate of $h$ needs to be computed; it coincides with the posterior mean given the observed data:
\begin{equation}\label{equ:post_est}
\hat{h} := \mathbb{E}_{\hat{\eta}} \left[h \vert Y_N \right] =\int h \frac{p(Y_N\vert h)p_{\hat{\eta}}(h)}{p_{\hat{\eta}}(Y_N)} dh
\end{equation}
In the SysId context, $h$ is typically modelled as a zero-mean Gaussian process (independent of the noise $e(t)$) with covariance $\mathbb{E}\left[h(t),h(s)\right]=\bar{K}_\eta(t,s)$ (aka kernel in the Machine Learning literature) \cite{GP-AC-GdN:11,ChenOL12}. Thanks to this asssumption, the marginal likelihood $p_\eta(Y_N)$ is Gaussian and the estimate \eqref{equ:post_est} is available in closed form.
\\
Furthermore, for simplicity the IIR model in \eqref{equ:sys} can be accurately approximated by a FIR model of order $n$, whenever $n$ is chosen large enough to catch the relevant components of the system dynamics. By collecting in $\mathbf{h}:=\left[h(1)\ \cdots\ h(n)\right]^\top\in\mathbb{R}^n$ the first $n$ impulse response coefficients, the following Gaussian prior can be defined:
\begin{align}
p_\eta(\mathbf{h})&\sim \mathcal{N}(0,K_\eta), \qquad \eta \in \Omega \subset \mathbb{R}^d,\ \ K_\eta \in \mathbb{R}^{n\times n}
\end{align}
The hyper-parameters $\eta$ can then be estimated by solving
\begin{align}
\hat{\eta} &= \arg\min_{\eta\in\Omega}\ - \ln p_\eta(Y_N)= \arg\min_{\eta\in\Omega}\ f_N(\eta)\label{equ:ml_max}\\
f_N(\eta) &= Y_N^\top \Sigma(\eta)^{-1} Y_N + \ln \det \Sigma(\eta)\label{equ:ml}\\
\Sigma(\eta)&= \Phi_N K_\eta \Phi_N^\top + \sigma^2 I_N
\end{align}
where $\Phi_N\in\mathbb{R}^{N\times n}$:
\begin{equation}\label{equ:phi}
\Phi_N := \begin{bmatrix}
u(0) & u(-1) & \cdots & u(-n+1)\\
\vdots & \ddots & \ddots &\vdots\\
u(N) & u(N-1) & \cdots & u(N-n+1)
\end{bmatrix} 
\end{equation}
In the batch setting we are considering  the quantities $u(-n+1), ...,u(0)$ can be either estimated or set to zero. Here, we follow the latter option. 
Once $\hat{\eta}$ has been computed, the corresponding minimum variance estimate is given by
\resizebox{1.02\linewidth}{!}{
  \begin{minipage}{\linewidth}
\begin{align}
\widehat{\mathbf{h}}  :&=\mathbb{E}_{\hat{\eta}}\left[\mathbf{h} \vert Y_N\right] = \arg\min_{\mathbf{h}^\in\mathbb{R}^n} \left(Y_N-\Phi_N\mathbf{h}\right)^\top \left(Y_N-\Phi_N\mathbf{h}\right) + \sigma^2 \mathbf{h}^\top K_{\hat{\eta}}^{-1}\mathbf{h}\nonumber \\ 
&=  (\Phi_N^\top \Phi_N + \sigma^2 K_{\hat{\eta}}^{-1})^{-1}\Phi_N^\top Y_N  \label{equ:h_hat}
\end{align}
\end{minipage}
}

\vspace{1mm}

\begin{rem}
The estimate $\widehat{\mathbf{h}}$ in \eqref{equ:h_hat} can be computed once a noise variance estimate $\hat{\sigma}^2$ is available. For this purpose, $\sigma^2$ can be treated as a hyper-parameter and estimated by solving \eqref{equ:ml_max} or it can be computed from a LS estimate of $\mathbf{h}$. In this work the latter option is adopted.
\end{rem}

\subsection{Online Approach} \label{subsec:bayes_onine}
\noindent We now adapt the batch technique described in Sec.~\ref{subsec:batch_approach} to the online setting outlined in Sec.~\ref{sec:problem_formulation}. At time $k+T$, when data $\mathcal{D}_{i+1}=\left\{u(t),y(t)\right\}_{t=iT+1}^{(i+1)T}$ are provided, the current impulse response estimate $\widehat{\mathbf{h}}^{(i)}$ is updated through formula \eqref{equ:h_hat}, once the data matrices are enlarged with the new data and a new hyper-parameter estimate $\hat{\eta}^{(i+1)}$ is computed. The data matrices are updated through the following recursions
\begin{align}
R^{(i+1)} &:= \Phi_{(i+1)T}^\top \Phi_{(i+1)T} = R^{(i)} + \left(\Phi_{iT+1}^{(i+1)T}\right)^\top \Phi_{iT+1}^{(i+1)T}\label{equ:r}\\
\widetilde{Y}^{(i+1)} &:= \Phi_{(i+1)T}^\top Y_{(i+1)T} =\widetilde{Y}^{(i)} + \left(\Phi_{iT+1}^{(i+1)T}\right)^\top Y_{iT+1}^{(i+1)T}\label{equ:y_tilde}\\
\xbar{Y}^{(i+1)} &:= Y_{(i+1)T}^\top Y_{(i+1)T}= \xbar{Y}^{(i)}  + \left(Y_{iT+1}^{(i+1)T}\right)^\top Y_{iT+1}^{(i+1)T}\label{equ:y_bar}
\end{align}
where $Y_{(i+1)T}=\left[y(1)\cdots y(iT+T)\right]^\top\in\mathbb{R}^{(i+1)T}$, $Y_{iT+1}^{(i+1)T} = \left[y(iT+1) \cdots y(iT+T)\right]$; $\Phi_i$ is defined as in \eqref{equ:phi} with $N$ replaced by $(i+1)T$, while $\Phi_{iT+1}^{(i+1)T}$ has the same structure of matrix \eqref{equ:phi} but it contains the data from $iT-n+1$ to $(i+1)T$.
The computational cost of \eqref{equ:r}-\eqref{equ:y_bar} is, $O(n^2T)$, $O(nT)$ and $O(T^2)$, respectively.\\
The minimization of $f_{(i+1)T}(\eta)$ in \eqref{equ:ml}, needed to determine $\hat{\eta}^{(i+1)}$, is typically performed through iterative routines, such as 1st or 2nd order optimization algorithms \cite{BonettiniCPSIAM2014} or the Expectation-Maximization (EM) algorithm \cite{bottegal2016robust,BOTTEGAL2015466}. Since these methods may require a large number of iterations before reaching convergence, they may be unsuited for online applications. We should recall that, when adopted for marginal likelihood optimization, each iteration of these algorithms has a computational complexity of $O(n^3)$, due to the objective function evaluation. Specifically, $f_{(i+1)T}(\eta)$ can be robustly evaluated as \cite{chen2013implementation}
\begin{align}
f_{(i+1)T}(\eta) = &((i+1)T-n)\ln \sigma^2 + 2\ln\vert S\vert \nonumber\\
&+\sigma^{-2}(\ \xbar{Y}^{(i+1)}- \| S^{-1}L^\top \widetilde{Y}^{(i+1)}\|_2^2\ ) \label{equ:eff_ml}
\end{align}
where $L$ and $S$ are Cholesky factors: $K_\eta =: LL^\top$ and $\sigma^2 I_n + L^\top R^{(i+1)} L =: SS^\top$ (whose computation is $O(n^3)$).\\ 
To tackle the real-time constraints, the approach proposed in \cite{RPPCECC2016} is adopted: $\hat{\eta}^{(i+1)}$ is computed by running just one iteration of a Scaled Gradient Projection (SGP) algorithm (a 1st order optimization method) applied to solve problem \eqref{equ:ml_max} \cite{BonettiniCPSIAM2014}. Algorithm \ref{alg:1step_grad} summarizes its implementation. Notice that it is initialized with the previous estimate $\hat{\eta}^{(i)}$ (obtained using the data $\bigcup_{l=1}^{i} \mathcal{D}_l$) which is likely to be close to a local optimum of the objective function $f_{iT}(\eta)\equiv f_{k}(\eta)$. If the number of new data $T << k$, it is reasonable to suppose that $\arg\min_{\eta\in\Omega} f_{iT} (\eta) \approx \arg\min_{\eta\in \Omega} f_{(i+1)T} (\eta)$. 
Therefore, by just performing one SGP iteration, $\hat{\eta}^{(i+1)}$ will be sufficiently close to a local optimum of $f_{(i+1)T}(\eta)$.

\begin{algorithm}
\caption{1-step Scaled Gradient Projection (SGP)}\label{alg:1step_grad}
\begin{algorithmic}[1]
\Statex{\textbf{Inputs:}} previous estimates $\{ \hat{\eta}^{(i)}, \hat{\eta}^{(i-1)}\}$, $\nabla f_{iT}(\hat{\eta}^{(i-1)})$, $R^{(i+1)}$, $\widetilde{Y}^{(i+1)}$, $\xbar{Y}^{(i+1)}$, $\hat{\sigma}^{(i+1)^2}$
\Statex Initialize: $c=10^{-4},\ \delta=0.4$
\State Compute $\nabla f_{(i+1)T}(\hat{\eta}^{(i)})$
\State $r^{(i-1)} \gets \hat{\eta}^{(i)} - \hat{\eta}^{(i-1)}$ \label{alg_step:rf}
\State $w^{(i-1)} \gets \nabla f_{(i+1)T}(\hat{\eta}^{(i)}) - \nabla f_{iT}(\hat{\eta}^{(i-1)})$ \label{alg_step:w}
\State Approximate the inverse Hessian of $f_{(i+1)T}(\hat{\eta}^{(i)})$ as $B^{(i)}=\alpha^{(i)}D^{(i)}$ (using the procedure outlined in \cite{BonettiniCPSIAM2014})\label{alg_step:inverse_H}
\State Project onto the feasible set:
\Statex $z\gets \Pi_{\Omega,D^{(i)}} (\ \hat{\eta}^{(i)} -B^{(i)}\nabla f_{(i+1)T}(\hat{\eta}^{(i)})\ )$\label{alg_step:proj}
\State $\Delta\hat{\eta}^{(i)} \gets z - \hat{\eta}^{(i)}$
\State $\nu \gets 1$
\If{$f_{(i+1)T}(\hat{\eta}^{(i)}+\nu \Delta \hat{\eta}^{(i)}) \leq f_{(i+1)T}(\hat{\eta}^{(i)})+ c \nu \nabla f_{(i+1)T}(\hat{\eta}^{(i)})^\top \Delta\hat{\eta}^{(i)}$}
\State Go to step 12
\Else
\State $\nu \gets \delta \nu$
\EndIf
\State $\hat{\eta}^{(i+1)} \gets \hat{\eta}^{(i)} + \nu \Delta \hat{\eta}^{(i)}$
\Statex \textbf{Output:} $\hat{\eta}^{(i+1)}$
\end{algorithmic}
\end{algorithm}

\noindent The key step in Algorithm \ref{alg:1step_grad}  is \ref{alg_step:inverse_H}, where the inverse Hessian is approximated as the product between the positive scalar $\alpha^{(i)}\in\mathbb{R}_+$ and the diagonal matrix $D^{(i)}\in\mathbb{R}^{d\times d}$.
$\alpha^{(i)}$ is chosen by alternating the so-called Barzilai-Borwein (BB) rules \cite{barzilai1988two}:
\begin{equation}
\alpha_1^{(i)} := \frac{r^{(i-1)^\top}r^{(i-1)}}{r^{(i-1)^\top}w^{(i-1)}}, \qquad 
\alpha_2^{(i)} :=  \frac{r^{(i-1)^\top}w^{(i-1)}}{w^{(i-1)^\top}w^{(i-1)}}\label{equ:bb}
\end{equation}
with $r^{(i-1)}$ and $w^{(i-1)}$ specified at steps \ref{alg_step:rf} and~\ref{alg_step:w} of Algorithm \ref{alg:1step_grad}. The definition of $D^{(i)}$ depends on the constraints set and on the objective function. The authors in \cite{BonettiniCPSIAM2014} exploit the following decomposition of $\nabla_\eta f_{(i+1)T}(\eta)$ (defined in \eqref{equ:ml}):
\begin{equation}\label{equ:grad_decomp}
\nabla_\eta f_{(i+1)T}(\eta) = V(\eta) - U(\eta), \quad V(\eta)>0, \  U(\eta)\geq 0
\end{equation}
to specify $D^{(i)}$.
\noindent We refer the interested reader to \cite{BonettiniCPSIAM2014} for further details.

\noindent The projection operator adopted at step \ref{alg_step:proj} of Algorithm \ref{alg:1step_grad} is
\begin{equation}
\Pi_{\Omega,D^{(i)}} (z) = \textstyle{\arg\min_{x\in\Omega}} (x-z)^\top D^{(i)^{-1}}(x-z)
\end{equation}

\begin{rem}
Besides SGP, in \cite{RPPCECC2016} other inverse Hessian approximations are investigated (e.g. the BFGS formula). In this work we only consider the SGP approximation, since it appears preferable to the others, according to the experiments we performed (both in the time-invariant and -variant domain). \cite{RPPCECC2016} also considers the EM algorithm as an alternative to 1st order optimization methods to solve problem \eqref{equ:ml_max}. Even if the results reported for EM in \cite{RPPCECC2016} are comparable to the ones achieved through SGP, the latter approach appears superior to EM in the time-varying setting we are considering.
\end{rem}

\subsection{Dealing with time-varying systems}\label{sec:time_var}
\noindent In this section we deal with the identification of time-varying systems: specifically, estimators have to be equipped with tools through which past data become less relevant for the current estimation. In the following we propose two routines which combine the ``online Bayesian estimation'' above sketched with the ability to ``forget'' past data.

\subsubsection{Fixed Forgetting Factor}
Following a classical practice in parametric SysId (see Sec.~\ref{sec:pem}), we introduce a forgetting factor $\bar{\gamma} \in (0,1]$ into the data we are provided in order to base the estimation mainly on the more recent data. Specifically, we assume that the first $k$ data are generated according to the following linear model:
\begin{equation}\label{equ:ff_model}
\bar{G}_k Y_k = \bar{G}_k \Phi_k \mathbf{h} + E, \ E= \left[e(1)...e(k)\right]^\top \sim \mathcal{N}(0,\sigma^2 I_k)
\end{equation}
where $\bar{G}_k \bar{G}_k =: \bar{\Gamma}_k$  and $\bar{\Gamma}_k := diag\left(\bar{\gamma}^{k-1}, \bar{\gamma}^{k-2}, ..., \bar{\gamma}^0\right)$. Therefore, when adopting the regularized regression criterion \eqref{equ:h_hat}, the estimate at time $k$ is computed as:
\begin{align}
\widehat{\mathbf{h}}_{\bar{\gamma}} &:= \arg\min_{\mathbf{h}\in\mathbb{R}^n}\sum_{t=1}^k \bar{\gamma}^{k-t} \left(y(t) - \Phi_t^t \mathbf{h}\right)^2 + \sigma^2 \mathbf{h}^\top K_{\hat{\eta}}^{-1} \mathbf{h} \label{equ:regul_probl_ff}\\
&= \arg\min_{\mathbf{h}\in\mathbb{R}^n}\left(Y_k - \Phi_k \mathbf{h}\right)^\top \bar{\Gamma}_k \left(Y_k - \Phi_k \mathbf{h}\right) + \sigma^2 \mathbf{h}^\top K_{\hat{\eta}}^{-1} \mathbf{h}\nonumber\\
&= (\Phi_k^\top \bar{\Gamma}_k \Phi_k + \sigma^2 K_{\hat{\eta}}^{-1})^{-1} \Phi_k^\top \bar{\Gamma}_k Y_k \label{equ:h_hat_ff}
\end{align}
Correspondingly, the hyper-parameters are estimated solving:
\begin{align}
\hat{\eta} &= \arg\min_{\eta\in\Omega}\ Y_k^\top \bar{G}_k \Sigma_{\bar{\gamma}}(\eta)^{-1} \bar{G}_k Y_k + \ln \det \Sigma_{\bar{\gamma}}(\eta)\label{equ:eta_hat_ff}\\
\Sigma_{\bar{\gamma}}(\eta)&= \bar{G}_k \Phi_k K_\eta \Phi_k^\top \bar{G}_k + \sigma^2 I_k
\end{align}
Algorithm \ref{alg:on_line_ff} illustrates the online implementation of the identification procedure based on equations \eqref{equ:h_hat_ff} and \eqref{equ:eta_hat_ff}. In particular, it assumes that at time $k$ the estimates $\widehat{\mathbf{h}}^{(i)}$ and $\hat{\eta}^{(i)}$ are available and they have been computed by solving, respectively, \eqref{equ:regul_probl_ff} and \eqref{equ:eta_hat_ff}; these estimates are then online updated after the new data $\mathcal{D}_{i+1}$ are provided. Once $\bar{\gamma}$ is chosen by the user, it is inserted in the data matrices $R_{\bar{\gamma}}^{(i+1)}:=\Phi_{(i+1)T}^\top \bar{\Gamma}_{(i+1)T}\Phi_{(i+1)T},\  \widetilde{Y}_{\bar{\gamma}^{(i+1)}}:=\Phi_{(i+1)T}^\top \bar{\Gamma}_{(i+1)T}Y_{(i+1)T}, \
\xbar{Y}_{\bar{\gamma}^{(i+1)}}:=Y_{(i+1)T}^\top \bar{\Gamma}_{(i+1)T}Y_{(i+1)T}$, updated at steps \ref{alg2_step:r}-\ref{alg2_step:yb} of the algorithm.

\begin{algorithm}
\caption{Online Bayesian SysId: Fixed Forgetting Factor}\label{alg:on_line_ff}
\begin{algorithmic}[1]
\Statex{\textbf{Inputs:}} forgetting factor $\bar{\gamma}$, previous estimates $\{ \hat{\eta}^{(i)}, \hat{\eta}^{(i-1)}\}$, previous data matrices $\{R_{\bar{\gamma}}^{(i)},\widetilde{Y}_{\bar{\gamma}}^{(i)},\xbar{Y}_{\bar{\gamma}}^{(i)}\}$, new data $\mathcal{D}_{i+1}=\left\{u(t),y(t)\right\}_{t=iT+1}^{(i+1)T}$
\State $R_{\bar{\gamma}}^{(i+1)} \gets \bar{\gamma}^T R_{\bar{\gamma}}^{(i)} + \left(\Phi_{iT+1}^{(i+1)T}\right)^\top \bar{\Gamma}_T\ \Phi_{iT+1}^{(i+1)T} $ \label{alg2_step:r}
\State $\widetilde{Y}_{\bar{\gamma}}^{(i+1)} \gets \gamma^T\widetilde{Y}_\gamma^{(i)} + \left(\Phi_{iT+1}^{(i+1)T}\right)^\top \bar{\Gamma}_T\ Y_{iT+1}^{(i+1)T}$ \label{alg2_step:yt}
\State $\xbar{Y}_{\bar{\gamma}}^{(i+1)} \gets \bar{\gamma}^T \xbar{Y}_{\bar{\gamma}}^{(i)}  + \left(Y_{iT+1}^{(i+1)T}\right)^\top \bar{\Gamma}_T\ Y_{iT+1}^{(i+1)T}$ \label{alg2_step:yb}
\State $\widehat{\mathbf{h}}_{LS}^{(i+1)} \gets R_{\bar{\gamma}}^{(i+1)^{-1}} \widetilde{Y}_{\bar{\gamma}}^{(i+1)}$ \label{alg2_step:ls}
\State {\footnotesize$\hat{\sigma}^{(i+1)^2} \gets \frac{1}{ (i+1)T - n} \left(\bar{Y}_{\bar{\gamma}}^{(i+1)}-2\widetilde{Y}_{\bar{\gamma}}^{(i+1)^\top}\widehat{\mathbf{h}}_{LS}^{(i+1)} + \widehat{\mathbf{h}}_{LS}^{(i+1)^\top}R_{\bar{\gamma}}^{(i+1)}\widehat{\mathbf{h}}_{LS}^{(i+1)} \right)$}
\State $\hat{\eta}^{(i+1)}\gets \arg\min_{\eta\in\Omega}\ f_{(i+1)T}(\eta)$ (use Algorithm \ref{alg:1step_grad})
\label{alg2_step:ml}
\State $\widehat{\mathbf{h}}^{(i+1)} \gets \left(R_{\bar{\gamma}}^{(i+1)} +\hat{\sigma}_{\bar{\gamma}}^{(i+1)^2} K_{\hat{\eta}^{(i+1)}}^{-1}\right)^{-1}\widetilde{Y}_{\bar{\gamma}}^{(i+1)}$\label{alg_step:h}
\Statex{\textbf{Output:}} $\widehat{\mathbf{h}}^{(i+1)}$, $\hat{\eta}^{(i+1)}$
\end{algorithmic}
\end{algorithm}

\subsubsection{Treating the Forgetting Factor as a Hyper-parameter}
The Bayesian framework provides the user with the possibility to treat the forgetting factor as a hyper-parameter and to estimate it by solving:
\begin{align}
\hat{\eta}, \hat{\gamma} &= {\textstyle\arg\min_{\eta\in\Omega, \gamma\in(0,1]}}\ f_k(\eta,\gamma)\label{equ:eta_hat_ff_hyper} \\
f_k(\eta, \gamma)&= Y_k^\top G_k \Sigma(\eta,\gamma)^{-1} G_k Y_k + \ln \det \Sigma(\eta,\gamma)\label{equ:ml_ff_hyper} \\
\Sigma(\eta,\gamma)&= G_k \Phi_k K_\eta \Phi_k^\top G_k + \sigma^2 I_k
\end{align}
where $G_k G_k =: \Gamma_k$  and $\Gamma_k := diag\left(\gamma^{k-1}, \gamma^{k-2}, ..., \gamma^0\right)$.
\begin{rem} Notice that the model \eqref{equ:ff_model} is equivalent to
$$
Y_k =  \Phi_k \mathbf{h} + E_{\bar{\gamma}}, \quad E_{\bar{\gamma}} = \left[e_{\bar{\gamma}}(1),...,e_{\bar{\gamma}}(k)\right]^\top \sim \mathcal{N}(0,\sigma^2 \bar{\Gamma}_k^{-1})
$$
Therefore, treating the forgetting factor as a hyper-parameter is equivalent to modeling the noise with a non-constant variance and to give to the diagonal entries of the covariance matrix an exponential decaying structure.
\end{rem}

The online implementation of this approach is detailed in Algorithm \ref{alg:on_line_ff_hyper}, where
\begin{equation}
R_{\hat{\pmb{\gamma}}}^{(i)} := \hat{\gamma}^{(i)} R_{\hat{\pmb{\gamma}}}^{(i-1)} + \left(\Phi_{(i-1)T+1}^{iT}\right)^\top \widehat{\Gamma}_T^{(i)} \Phi_{(i-1)T+1}^{iT} 
\end{equation}
with $\widehat{\Gamma}_T^{(i)} = diag((\hat{\gamma}^{(i)})^{T-1},.. ,(\hat{\gamma}^{(i)})^{0})$. $\widetilde{Y}_{\hat{\pmb{\gamma}}}^{(i)}$ and $\xbar{Y}_{\hat{\pmb{\gamma}}}^{(i)}$ are analogously defined.\\
We should stress that the objective function in \eqref{equ:ml_ff_hyper} does not admit the decomposition \eqref{equ:grad_decomp}; we have
\begin{equation*}
\frac{\partial f_k(\eta,\gamma)}{\partial \gamma} = V(\eta,\gamma) + U(\eta,\gamma), \quad V(\eta,\gamma) >0, \ U(\eta,\gamma) \geq 0 
\end{equation*}
Thus, when $\gamma$ is treated as an hyper-parameter, Algorithm \ref{alg:1step_grad} is run setting $D^{(i)}=I_d$ at step \ref{alg_step:inverse_H}; $\alpha^{(i)}$ is still determined alternating the BB rules \eqref{equ:bb}.

%

\begin{algorithm}
\caption{Online Bayesian SysId: Forgetting Factor as a hyper-parameter}\label{alg:on_line_ff_hyper}
\begin{algorithmic}[1]
\Statex{\textbf{Inputs:}} previous estimates $\{ \hat{\eta}^{(i)}, \hat{\eta}^{(i-1)}, \hat{\gamma}^{(i)}, \hat{\gamma}^{(i-1)}\}$, previous data matrices $\{R_{\hat{\pmb{\gamma}}}^{(i)},\widetilde{Y}_{\hat{\pmb{\gamma}}}^{(i)},\xbar{Y}_{\hat{\pmb{\gamma}}}^{(i)}\}$, new data $\mathcal{D}_{i+1}=\left\{u(t),y(t)\right\}_{t=iT+1}^{(i+1)T}$
\State $R_{\gamma}^{(i+1)} \gets \gamma^T R_{\hat{\pmb{\gamma}}}^{(i)} + \left(\Phi_{iT+1}^{(i+1)T}\right)^\top \Gamma_T \ \Phi_{iT+1}^{(i+1)T} $ \label{alg3_step:r}
\State $\widetilde{Y}^{(i+1)}_{\gamma} \gets \gamma^T \widetilde{Y}_{\hat{\pmb{\gamma}}}^{(i)} + \left(\Phi_{iT+1}^{(i+1)T}\right)^\top \Gamma_T \ Y_{iT+1}^{(i+1)T}$ \label{alg3_step:yt}
\State $\xbar{Y}^{(i+1)}_\gamma \gets \gamma^T \xbar{Y}_{\hat{\pmb{\gamma}}}^{(i)}  + \left(Y_{iT+1}^{(i+1)T}\right)^\top \Gamma_T \ Y_{iT+1}^{(i+1)T}$ \label{alg3_step:yb}
\State $\widehat{\mathbf{h}}_{LS}^{(i+1)} \gets (R_{\hat{\pmb{\gamma}}}^{(i)})^{-1} \widetilde{Y}_{\hat{\pmb{\gamma}}}^{(i)}$ \label{alg3_step:ls}
\State {\footnotesize$\hat{\sigma}^{2^{(i+1)}} \gets \frac{1}{ (i+1)T - n} \left(\xbar{Y}_{\hat{\pmb{\gamma}}}^{(i)}-
2(\widetilde{Y}_{\hat{\pmb{\gamma}}}^{(i)})^\top\ \widehat{\mathbf{h}}_{LS}^{(i+1)} + (\widehat{\mathbf{h}}_{LS}^{(i+1)})^\top R_{\hat{\pmb{\gamma}}}^{(i)} \widehat{\mathbf{h}}_{LS}^{(i+1)} \right)$}
\State $\hat{\eta}^{(i+1)}, \hat{\gamma}^{(i+1)} \gets \arg\min_{\eta\in\Omega, \gamma\in(0,1]}\ f_{(i+1)T}(\eta,\gamma)$ \Statex (use Algorithm \ref{alg:1step_grad})
\label{alg3_step:ml}
\State $\widehat{\mathbf{h}}^{(i+1)} \gets \left(R_{\hat{\pmb{\gamma}}}^{(i+1)} +\hat{\sigma}^{2^{(i+1)}}(\hat{\gamma}^{(i+1)})\ K_{\hat{\eta}^{(i+1)}}^{-1}\right)^{-1}
\widetilde{Y}_{\hat{\pmb{\gamma}}}^{(i+1)}$\label{alg_step:h}
\Statex{\textbf{Output:}} $\widehat{\mathbf{h}}^{(i+1)}$, $\hat{\eta}^{(i+1)}$
\end{algorithmic}
\end{algorithm}

\section{Experimental Results}\label{sec:experiment}
\noindent In this section we test the online algorithms for parametric and Bayesian SysId described in Sec.~\ref{sec:pem} and \ref{sec:bayes}. Their performance are compared through a Monte-Carlo study over 200 time-varying systems.

\subsection{Data}
\noindent 200 datasets consisting of 3000 input-output measurement pairs are generated. Each of them is created as follows: the first 1000 data are produced by a system contained in the data-bank D4 (used in \cite{TC-MA-LL-AC-GP:14}), while the remaining 2000 data are generated by perturbing the D4-system with two additional poles and zeros. These  are chosen such that the order of the D4-system changes, thus creating a switch on the data generating system at time $k=1001$. 
\\
The data-bank D4 consists of 30th order random SISO dicrete-time systems having all the poles inside a circle of radius 0.95. These systems are simulated with a  unit variance  band-limited Gaussian signal with normalized band $[0,0.8]$. A zero mean white Gaussian noise,  with  variance adjusted so that the Signal to Noise Ration (SNR) is always equal to 1, is then added to the output data.

\subsection{Estimators}
\noindent The parametric estimators are computed with the \verb!roe! Matlab routine, using the BIC criterion for the model complexity selection. In the following this estimator will be denoted as \textit{PEM BIC}. Furthermore, as a benchmark we introduce the parametric oracle estimator, called \textit{PEM OR}, which selects the model complexity by choosing the model that gives the best fit to the impulse response of the true system. The order selection is performed every time a new dataset becomes available: multiple models with orders ranging from 1 to 20 are estimated and the order selection is performed according to the two criteria above-described.
\\
For what regards the methods relying on Bayesian inference we adopt a zero-mean Gaussian prior with a covariance matrix (kernel) given by the so-called TC-kernel:
\begin{equation}
K_\eta^{TC}(k,j) = \lambda \min(\beta^{k},\beta^{j}), \qquad \eta = [ \lambda,\ \beta]
\end{equation}
with $\Omega = \{(\lambda,\beta): \lambda \geq 0, 0 \leq \beta \leq 1\}$ \cite{ChenOL12}. The length $n$ of the estimated impulse responses is set to 100. In the following, we will use the acronym \textit{TC} to denote these methods. Furthermore, the notation \textit{OPT} will refer to the standard Bayesian procedure, in which the SGP algorithm adopted to optimize the marginal likelihood $f_k(\eta)$ is run until the relative change in $f_k(\eta)$ is less than $10^{-9}$.
From here on, the online counterpart (illustrated in Sec.~\ref{sec:bayes}) will be referred to as the \textit{1-step ML}.
We will also use the acronyms \textit{TC FF} when a fixed forgetting factor is adopted, \textit{TC est FF} when the forgetting factor is estimated as a hyper-parameter.
\\
For each Monte Carlo run, the identification algorithms are initialized using the first batch of data $\mathcal{D}_{init}=\left\{u(t),y(t)\right\}_{t=1}^{300}$.
After this initial step, the estimators are updated every $T=10$ time steps, when new data $\mathcal{D}_{i+1} = \left\{u(t),y(t)\right\}_{t=iT}^{(i+1)T}$ are provided. The forgetting factor in the \textit{TC FF} and \textit{PEM} methods is set to 0.998, while its estimation in \textit{TC est FF} method is initialized with 0.995.

\subsection{Performance}
\begin{figure}[hbtp]
\centering
\includegraphics[width=.9\columnwidth]{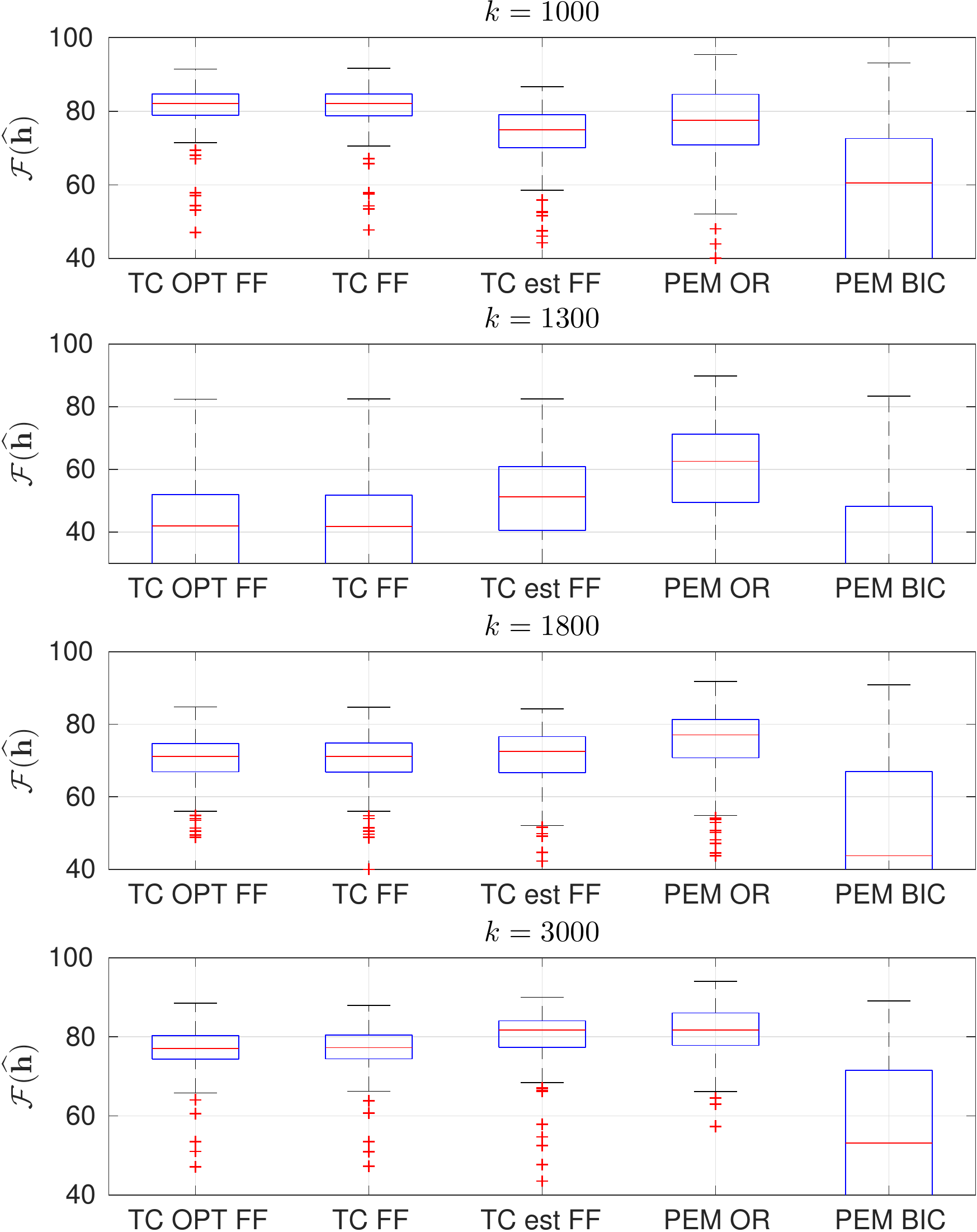}
\caption{Fit $\mathcal{F}(\widehat{\mathbf{h}})$ achieved at four time instants $k$ (corresponding to the number of data available for the estimation).}
\label{fig:fit}
\end{figure}

\noindent The purpose of the experiments is twofold. First, we will compare the two routines we have proposed in Sec.~\ref{sec:time_var} to explicitly deal with time-varying systems. Second, we will compare the parametric and the Bayesian identification approaches while dealing with time-varying systems.\\
As a first comparison, we evaluate the adherence of the estimated impulse response $\widehat{\mathbf{h}}$ to the true one $\mathbf{h}$, measured~as
\begin{equation}
\label{eq:fit_imp_resp}
\mathcal{F}(\widehat{\mathbf{h}})= 100 \cdot \Big(1- \frac{\Vert \mathbf{h} - \widehat{\mathbf{h}} \Vert_2}{\Vert \mathbf{h} \Vert_2}\Big)
\end{equation}
Figure \ref{fig:fit} reports the values of $\mathcal{F}(\widehat{\mathbf{h}})$ at four time instants.

\noindent 
It is interesting to note that immediately before the change in the data generating system ($k = 1000$) the \textit{TC} methods slightly outperform the ideal parametric estimator \textit{PEM OR}.
%
After the switch (occurring at $k=1001$), among regularization/Bayesian routines \textit{TC est FF} recovers the fit performance a bit faster than \textit{TC FF}; even at regime it outperforms the latter because it can choose forgetting factor values that retain a larger amount of data.
\\
We also observe how the \textit{1-step ML} procedures and the corresponding \textit{OPT} routines provide analogous performance at each time step $k$, validating the method we propose for online estimation and confirming the results in \cite{RPPCECC2016}.
\\
The unrealistic \textit{PEM OR} represents the reference on the achievable performance of the PEM estimators; it outperforms \textit{TC} methods in the transient after the switch, while it has comparable performance at regime. Instead, the recursive \textit{PEM BIC} estimator performs very poorly.


\begin{table}
\centering
\begin{tabularx}{\columnwidth}{|X|X|X|X|X|X|X|}
\hline 
 & \multicolumn{3}{c|}{TC}  & \multicolumn{2}{c|}{PEM}  
\\ 
 & OPT FF & FF & est FF & OR & BIC \\ 
\hline 
mean & 6.70  &       0.44  &       5.45  &   18.44  &      18.44  \\\hline
std & 1.28  &       0.03  &       0.67  &     0.69  &       0.69  \\
\hline 
\end{tabularx} 
\caption{Computational cumulative time after data $\mathcal{D} =\left\{u(t),y(t)\right\}_{t=1}^{3000}$ are used: mean and std over 200 datasets.}\label{tab:cumulative_time_TC_PEM}
\vspace{-4mm}
\end{table}

\noindent As a second comparison, Table \ref{tab:cumulative_time_TC_PEM} reports the computational cumulative time of the proposed algorithms in terms of mean and standard deviation after the estimators are fed with all the data $\mathcal{D} =\left\{u(t),y(t)\right\}_{t=1}^{3000}$. 
The \textit{1-step ML} methods are one order of magnitude faster than the corresponding \textit{OPT} ones. The \textit{TC est FF} estimator is  slower than \textit{TC FF}: this should be a consequence of having set $D^{(i)}=I_d$ in Algorithm \ref{alg:1step_grad}. On the other hand the RPEM estimators are three times slower than the \textit{OPT} ones, thus appearing not particularly appealing for online applications. 
\section{Conclusion and Future Work}\label{sec:conclusion}
\noindent We have adopted recently developed SysId techniques relying on the Gaussian processes and Bayesian inference to the identification of time-varying systems. Specifically, we have focused on an online setting by assuming that new data become available at predefined time instants. To tackle the real-time constraints we have modified the standard Bayesian procedure: hyper-parameters are estimated by performing only one gradient step in the corresponding marginal likelihood optimization problem. 
In order to cope with the time-varying nature of the systems to be identified, we propose two approaches, based on the use of a forgetting factor. One of them treats the forgetting factor as a constant, while the other estimates it as a hyper-parameter of the Bayesian inference procedure.\\
We believe that the preliminary investigation performed in this work may pave the way for further research in this topic. A future research direction could consider the recursive update of the Bayesian estimate, resembling the one which is available for parametric techniques.

\bibliographystyle{IEEEtran}
\bibliography{References}

\end{document}